\documentclass[journal=jacsat,manuscript=article,layout=twocolumn]{achemso}

\usepackage[version=3]{mhchem} 
\usepackage[T1]{fontenc}       
\usepackage{graphicx,xcolor,charter,mathtools}
\graphicspath{{images/}}

\usepackage[english]{babel}
\usepackage[utf8]{inputenc}
\usepackage[colorinlistoftodos, color=green!40, prependcaption]{todonotes}
\usepackage{amsthm}
\usepackage{mathtools}
\usepackage{physics}
\usepackage{xcolor}
\usepackage{graphicx}
\SectionNumbersOn

\usepackage{tabu}
\usepackage{xcolor}
\usepackage{tabularx}
\usepackage{booktabs}
 
\usepackage[pdftex, pdftitle={Article}, pdfauthor={Author}]{hyperref} 
\title{Accessible Chemical Space for Metal Nitride Perovskites}

\author{Bastien F. Grosso}
\email{b.grosso@ucl.ac.uk}
\affiliation{Department of Chemistry, University College London, London, United Kingdom}
\author{Daniel W. Davies}
\affiliation{Department of Chemistry, University College London, London, United Kingdom}
\author{Bonan Zhu}
\affiliation{Department of Chemistry, University College London, London, United Kingdom}
\author{Aron Walsh}
\affiliation{Department of Materials, Imperial College London, London, United Kingdom}
\author{David O. Scanlon}
\email{d.scanlon@ucl.ac.uk}
\affiliation{Department of Chemistry, University College London, London, United Kingdom}

\begin{document}

\begin{abstract}
Building on the extensive exploration of metal oxide and metal halide perovskites, metal nitride perovskites represent a largely unexplored class of materials. We report a multi-tier computational screening of this chemical space. From a pool of 3660 ABN$_3$ compositions covering I-VIII, II-VII, III-VI and IV-V oxidation state combinations, 279 are predicted to be chemically feasible. The ground-state structures of the 25 most promising candidate compositions were explored through enumeration over octahedral tilt systems and global optimisation. We predict 12 dynamically and thermodynamically stable nitride perovskite materials, including YMoN$_3$, YWN$_3$, ZrTaN$_3$, and LaMoN$_3$. These feature significant electric polarisation and low predicted switching electric field, showing similarities with metal oxide perovskites and making them attractive for ferroelectric memory devices.

\end{abstract}


\maketitle

\section{Introduction}
In the past decades, nitride-based materials have paved the way to new technological paradigms with applications ranging from hard coatings (TiN), light-emitting diodes and high electron mobility transistors (GaN) to superconductors and high-power electronics (AlN).\cite{Qiang/Thumer/Reiners:1998,Seong/Han/Amano:2013,Lin/Chen/Lee:2020} The wide variety of properties emerging from nitride materials are directly related to the unique combination of high electronegativity and strong bonding character of the nitrogen atom itself. While binary nitrides unravel stunning properties, more can be expected from compositions and structures with increasing complexity. Although ternaries nitrides are expected to enhance the structure-property tunability further, very few systems have been identified up-to-date, with more than ten times fewer ternary nitrides known than ternary oxides.\cite{Sun/Bartel/Arca:2019} 

One of the most prolific systems for fine-tuning the structure-properties relationship is arguably the ABX$_3$ perovskite structure, where A and B are cations, and X is an anion bonding to A and B. In the ideal form, the A cations sit at the corners of a cube containing in its centre a BX$_6$ octahedron. This crystal structure can accommodate a wide range of distortions that, on the one hand, facilitate the growth of these materials as thin films, ideal for microelectronics, and on the other hand, provide an excellent playground for engineering new functionalities. The most famous families of perovskites are the halides, mainly known for photovoltaic applications, and the oxides, known for ferroelectricity and multiferroicity, among other properties. Recently, there has been a strong interest in discovering ternary nitrides with the perovskite structure, motivated by the hope to integrate them with the existing nitride-based semiconductor devices. Indeed, while oxide and halide perovskites have benefited from decades of intense research, they are hardly integrated into the current semiconductor technology. \cite{Moghadam/Xiao/Ahmadi:2017,Lin/Hong/Wood:2001} Therefore, the discovery of functional nitride perovskite materials would open the door to new standards in microelectronics.

Discovering new stable materials is extremely challenging. On the one hand, one can create an almost infinite number of hypothetical compounds by combining the elements in the periodic table. On the other hand, materials can form only if a complex energetic balance between electrons and nuclei is found. Therefore, trial and error approaches, both experimental and computational, are not viable for a comprehensive exploration of compositions. Hence, more modern methods, such as high throughput computational materials design, are nowadays preferred. This technique consists of screening many candidates based on criteria, sometimes complemented by intelligently interrogating existing database of materials,\cite{Curtarolo/Hart/Buogiorno:2013} to extract an affordable subset of candidates to be further investigated.\cite{Frey/Grosso/Fandre:2022} First-principles calculations, such as density-functional theory (DFT), are often employed to determine their stability and evaluate their properties. One would ideally use global structural prediction methods for crystal structure prediction,\cite{Oganov_crystStruct:2011,Wales_enLand:2004} to identify the ground state structure by searching for the configuration that globally minimises the energy. However, these techniques are computationally too demanding for high throughput studies and require further compromises, such as lowering the precision of the calculations or selecting a maximum size of the unit cell (i.e. a maximum number of atoms). A computationally more affordable and widely used approach relies on utilising prototype structures already identified for other materials, making the strong assumption that similar materials adopt similar structures.

Recently, several theoretical studies devoted to the discovery of potential nitride perovskites have explored a wide range of compositions selected either based on structural considerations such as Goldschmidt-like criteria,\cite{Sherbondy/Smaha/Bartel:2022,Ha/Lee/Giustino:2022} on the energy above the energy hull of the perfectly cubic perovskite structure \cite{SarmientoPerez/Cerqueira/Korbel:2015} or by fixing the B cation to certain elements and screening a series of elements for the A cation such as the lanthanides.\cite{Sherbondy/Smaha/Bartel:2022,FloresLivas/SarmientoPerez/Botti:2019} While having different starting points, these studies have in common that after selecting the candidate compositions, their stability and properties are calculated using prototype perovskite structures.  

In the current study, we propose a different approach to overcome some of the limitations of previous studies. We use chemical and electrostatic considerations (as opposed to structural) in the first place to filter candidates that are likely to exist. We then predict their structure starting from general octahedral tilts distortions imposed on top of the perovskite structure and carry out, in parallel, a crystal structure prediction to challenge the likelihood of our predicted nitrides ternaries to adopt a perovskite-like structure. Finally, after selecting the most favourable candidates, we use phonon-mapping to explore the potential energy surface of each material, identify its ground state and calculate its properties. Our approach is the most systematic screening procedure ever applied to nitride perovskites and yields 12 new dynamically and thermodynamically stable nitride perovskites, potentially synthesisable.

\section{Results and Discussion} 
\subsection{First Screening of compositions} \label{sec:first_screening}
The stability of a material adopting the perovskite structure (ABX$_3$) requires charge neutrality and electrostatic stability. These criteria are fulfilled if $q_A + q_B = 3q_X$, and $q_A \leq q_B$.\cite{Caetano/Butler/Walsh:2016} For oxide perovskites, this translates to three possible oxidation states for the cations: $A^{+1}B^{+5}$, $A^{+2}B^{+4}$ and $A^{+3}B^{+3}$. For nitride perovskites, assuming that nitrogen adopts its common oxidation state ($N^{-3}$), the A and B cations are constrained to oxidation states of $A^{+1}B^{+8}$, $A^{+2}B^{+7}$, $A^{+3}B^{+6}$ or $A^{+4}B^{+5}$. While the number of combinations is higher for nitrides, the high oxidation state of the B cation is unlikely to be stable for a wide range of compositions. 

\begin{figure}[H]
\includegraphics[width=\columnwidth]{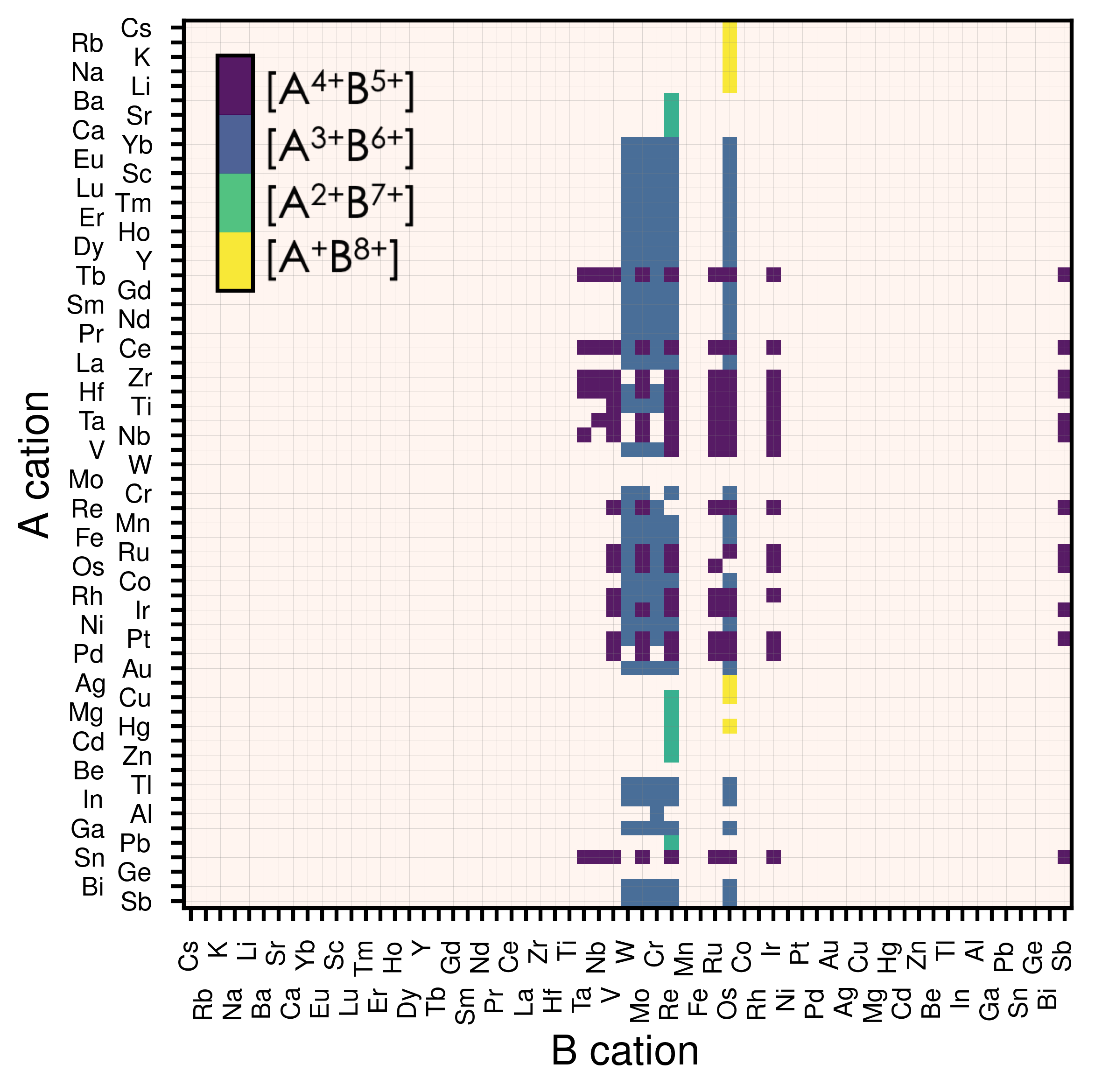}
\centering
\caption{Map of possible ABN$_3$ perovskite compositions. The A cations are displayed on the vertical axis and the B cations on the horizontal one. The oxidation states for the 279 candidates initially filtered are displayed by the four different colours, each representing an allowed combination of oxidation states. If a candidate composition has more than one combination of oxidation states, only that with the lowest oxidation state on the B-site is displayed. We use the Pettifor scale to order the atoms.\cite{Pettifor:1984}}
\label{fig_oxidstates}
\end{figure}

Thus, we start our search of A and B cations by considering combinations of 61 metallic elements (up to Bi), resulting in 3,660 candidate chemical systems. We use the SMACT \cite{Davies/Butler/Isayev:2018} package to filter out the compositions that do not allow charge neutrality and thus obtain 1,864 possible systems. To increase the likelihood of synthesising the selected candidates, we retain only those whose elements exhibit oxidation states observed in at least 5\% of the compounds reported in ICSD for each given element, which reduces the candidates to 374. \cite{ICSD_Levin} This is done using the list of oxidation states compiled in Ref. \cite{Ding/Kumagai/Oba:2020}. 
Consecutively, we impose that the cation on the B-site has a smaller or equal Shannon radius than that of the A cation. This last criterion and electrostatic stability maximise the chances of the perovskite structure holding together. Note that other more elaborated criteria, such as the tolerance factor or Goldschmidt rule,\cite{Goldschmidt:1926} exist for oxide and fluoride perovskites. Nevertheless, those are not well-defined for other families of materials with less ionic bonding, such as halide perovskites or nitride perovskites, in which the nitrogen radius is poorly defined.\cite{Travis/Glover/Bronstein:2016} Such criteria also fail for nitride perovskites simply because the size of the N ion is not well defined. Finally, we remove duplicate compositions (same A and B but with different oxidation states) and are left with 279 candidates, presented in Fig. \ref{fig_oxidstates}. We notice from Fig. \ref{fig_oxidstates} that most of our candidates adopt either $+5$ or $+6$ oxidation state on the B-site. This directly comes from the fact that we only considered realistic oxidation states by crosschecking them with oxidation states in synthesised materials.

\subsection{On the Importance of Tilts} \label{sec:tilts}

As mentioned earlier, one of the major interests of the perovskite structure is its ability to accommodate several distortions and rotations of the octahedra. The latter were classified and encoded in the well-known Glazer notation, resulting in 15 different tilts, each leading to a different space group.\cite{Glazer:1972}

We take advantage of the general character of the Glazer tilts and study their impact on the energy of each of our 279 candidates. We start with a volume relaxation of each material in the cubic phase ($a^0a^0a^0$, in Glazer notation). We then apply all 15 tilts to our 279 compounds and allow complete relaxation of the ions and the lattice vectors. The tilts are chosen to have the same amplitude for each combination but with the increasing angle in increments of $0.05$ rad, thus taking values of $0.1, 0.15$ and $0.2$ rad when $a,b$ and $c$ tilt angles all differ. 

\begin{figure*}[t]\
\includegraphics[width=\linewidth]{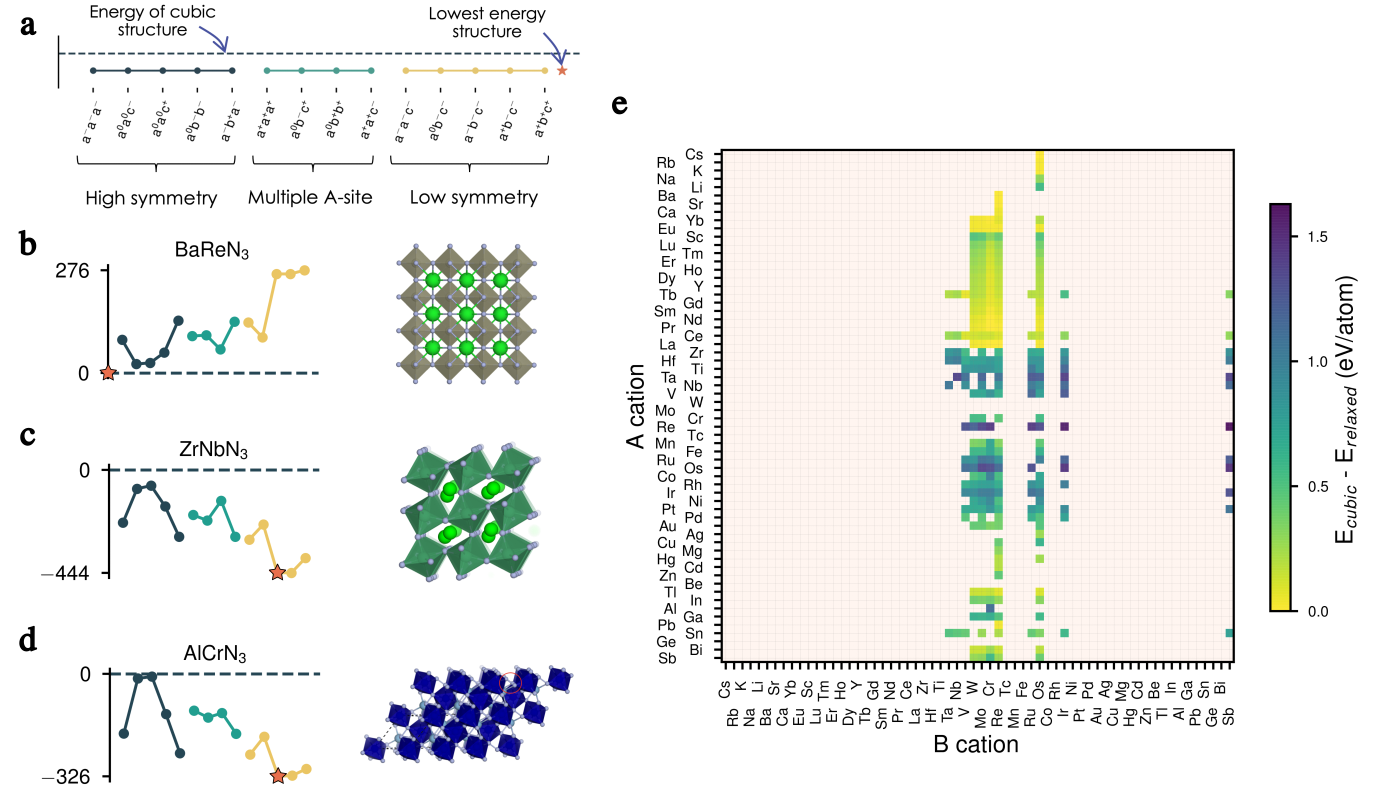}
\centering
\caption{Effect of octahedral tilts. 
(a) Legend for the tilt patterns with the cubic energy reference given as a dashed line. A star indicates the lowest energy structure.
(b) Unstable tilts for BaReN$_3$ (left) and the corresponding lowest-energy cubic structure (right). 
(c) Stable tilts for ZrNbN$_3$ (left) and corresponding lowest-energy perovskite structure (right). 
(d) Tilt energies for AlCrN$_3$ (left) and the corresponding lowest-energy non-perovskite structure (right). 
(e) Energy difference between the cubic and fully relaxed lowest-energy structure for all 279 candidates.
Note that all energies presented in (b-d) are obtained after only one loop of electronic relaxation, whereas in (e) each structure was fully relaxed (ions and lattice).}
\label{fig_tilts}
\end{figure*}\
We present in Fig. \ref{fig_tilts}(a-d) the effect of tilts on the energy for three of our candidates before full relaxation, displaying three different scenarios. In the first example, all tilts imposed on the cubic structure of BaReN$_3$ result in higher-energy structures and, therefore, are not favourable for this material (Fig. \ref{fig_tilts}b). In the second example, ZrNbN$_3$ lowers its energy for all tilts imposed, with the $a^-b^+a^-$ being the lowest-energy distortion pattern, resulting in a perovskite-like structure (Fig. \ref{fig_tilts}c). Finally, for AlCrN$_3$ some tilt patterns have a similar energy to the non-tilted structure, and others lower the energy, with $a^0b^-c^-$ being the lowest-energy tilt pattern but resulting in a structure not respecting the perovskite criteria \cite{Breternitz/Schorr:2018} (Fig. \ref{fig_tilts}d).

We repeat the same analysis for all candidates and present in Fig. \ref{fig_tilts}e the energy differences between the cubic and the fully relaxed lowest-energy structures. From this data, it is evident that the tilts play a significant role in bringing stability and can, therefore, not be omitted when evaluating the thermodynamic equilibrium of these materials. In other words, filtering candidates based on the cubic perovskite energy could inevitably discard promising candidates. The complete data set presenting the energies of all 15 tilted structures before and after complete relaxation for all candidates can be found in Fig. S1-S8 in the supplementary information. 

\begin{figure}[t]
\includegraphics[width=\columnwidth]{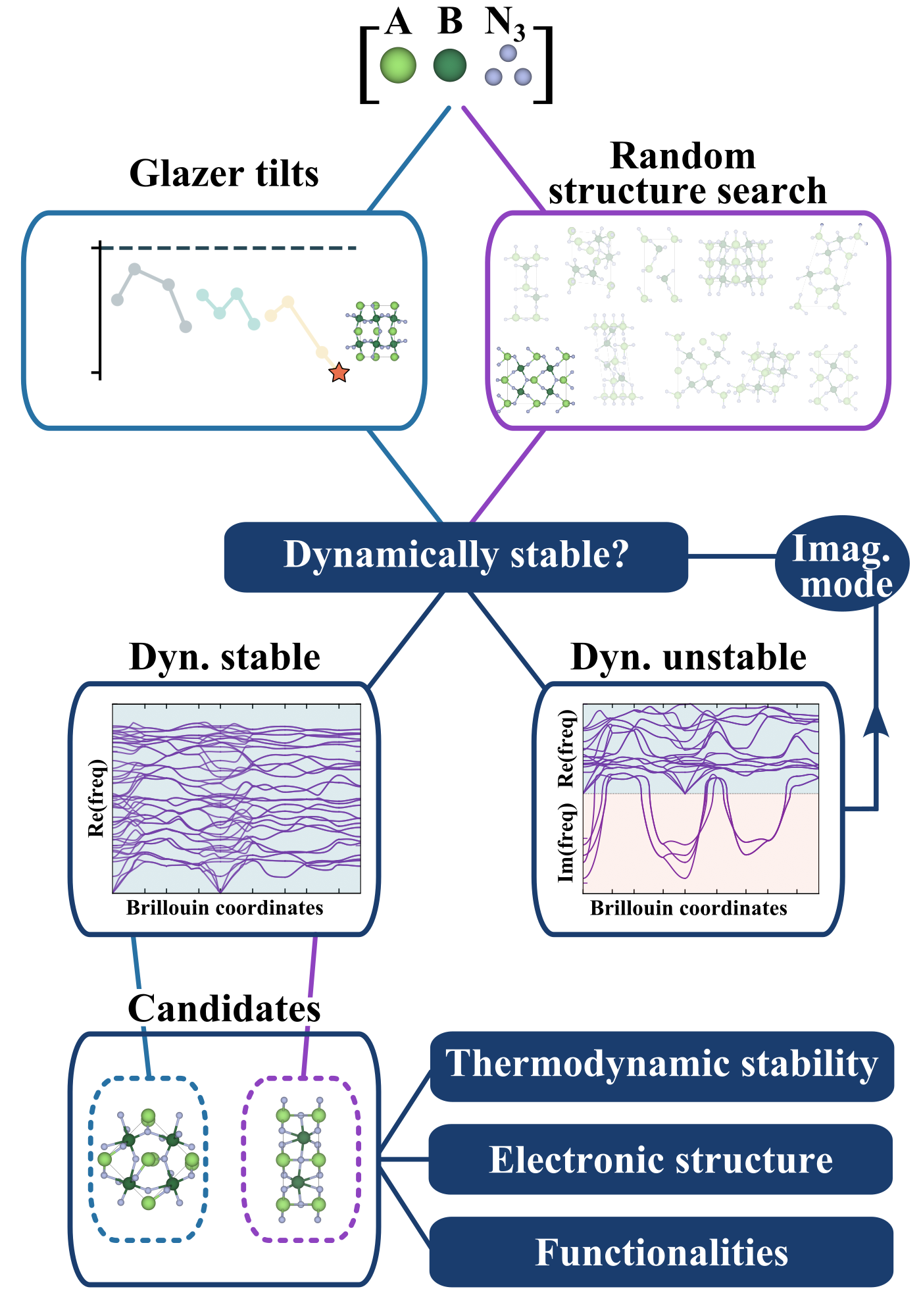}
\centering
\caption{Workflow to identify stable structures. We select the lowest-energy tilted structure for each candidate composition and the lowest-energy random structure. Dynamical stability is tested for each structure through an iterative process based on phonon frequencies (see Methods). Thermodynamic stability, electronic analysis, and the evaluation of functionalities are then performed.}
\label{fig_structuralSearch}
\end{figure}

\subsection{Second Screening of compositions} \label{sec:second_screening}

Even though all 279 candidates identified so far are chemically sensible and could, in principle, be stabilised, we decided to reduce this number to a computationally more affordable one by imposing further restrictions. While almost all metals are allowed on the A-site (Fig. \ref{fig_oxidstates}), only 11 elements are stable on the B-site in oxidation states between $+5$ and $+8$: Ta, Nb, V, W, Mo, Cr, Re, Ru, Os, Ir and Sb. Among them, the most significant phase space is composed of materials having $A^{+3}B^{+6}$ or $A^{+4}B^{+5}$ oxidation states. For ease, we limit our candidates to the compounds containing $B^{+5}$ and $B^{+6}$ closed shells and exclude lanthanides (except La) to avoid the challenging treatment of the highly-localised $f$ electrons due to their self-interaction errors, inherent from semi-local DFT.\cite{Cohen/Mori/Yang:2011} With these final constraints, we obtain 25 candidate compositions presented later when discussing thermodynamic stability.

\subsection{Structure prediction and dynamical stability} \label{sec:structure_dynamic}

So far, we have identified 25 nitride perovskite candidates and obtained for each of them the lowest-energy tilted structure. Before considering the possibility of synthesising these materials, we need to ensure that the crystal structures found are dynamically stable, equivalent to having no imaginary vibrational frequency over the whole Brillouin zone. While we assess dynamical stability based on the temperature-independent phonon dispersion, it is worth mentioning that it approximates the "true" stability. Indeed, phonon frequencies are, in principle, temperature and pressure dependent. Thus, a change in thermodynamic conditions could, in principle, stabilise a crystal structure found unstable at zero temperature.\cite{Tolborg/Klarbring/Ganose:2022}

 In Fig. \ref{fig_structuralSearch}, we present the workflow to find the crystal structures. Starting from the lowest-energy tilted structure found for each candidate, we compute phonon frequencies using the procedure described in the Methods section. In parallel, we do a random structure search using AIRSS\cite{AIRSS1:Pickard/Needs:2006,AIRSS2:Pickard/Needs:2011} (see Methods) and check the dynamical stability for all random structures. Note that we will refer from now on to the first structure as "Glazer", noted (G), and to the second one as "AIRSS", noted (A). 
 
 The objective of this complementary approach to identify the ground state is twofold. Firstly, starting from random configurations, as opposed to perovskite-like arrangements, allows exploring other sensible structures without biasing the initial point and therefore challenges the perovskite structure. Secondly, by further exploring the phase space of the materials, we can compute the energy hull more accurately and better estimate the formation energy and thermodynamic stability.

\begin{figure*}[t]\
\includegraphics[width=0.8\linewidth]{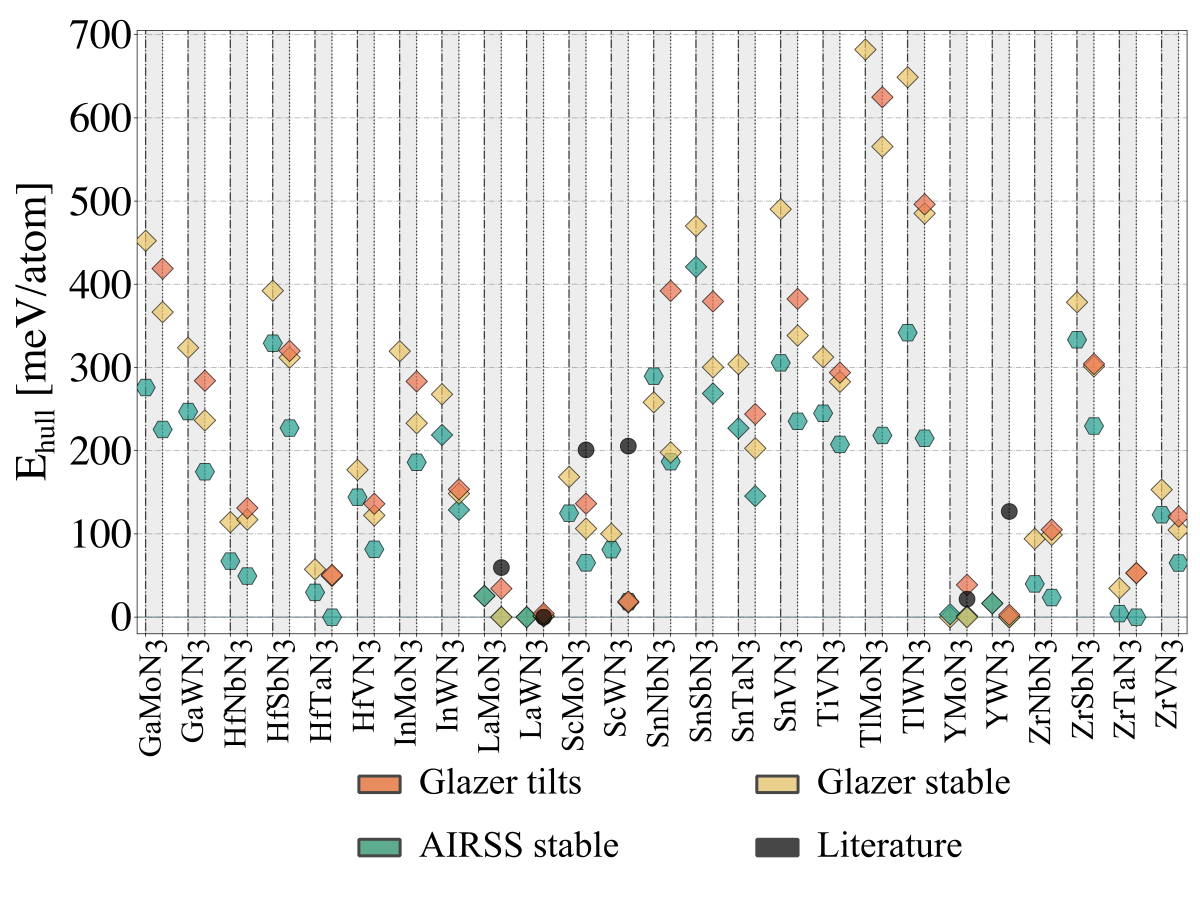}
\centering
\caption{Thermodynamic (energy above the hull) analysis for all candidate nitride perovskites. 
Each grey column corresponds to a material, with the HSE06 data plotted on the left (dash-dotted vertical lines) and the PBEsol data on the right (dotted vertical lines). A diamond indicates materials adopting a perovskite-like structure. The 'Glazer tilts' data corresponds to the lowest-energy tilted structure, whose energy was plotted in Fig. \ref{fig_tilts}e. The 'Glazer stable' and 'AIRSS stable' correspond to the dynamically stable structure obtained following the workflow presented in Fig. \ref{fig_structuralSearch}, starting from the Glazer and AIRSS structures, respectively. Note that InMoN$_3$ and TlMoN$_3$ AIRSS structures were not found to be dynamically stable, and only the PBEsol energy of the initial structure is reported; both structures are non-perovskite. The distance to the hull for the hybrid calculations can be found in Table \ref{tab_abc_E_P}.}
\label{fig_Thermodyn}
\end{figure*}

\subsection{Thermodynamic stability} \label{sec:thermodyn}
To evaluate our compounds' thermodynamic stability, we compute the distance to the energy convex hull using PYMATGEN.\cite{PingOng/DavidsonRichards/Jain:2013} The hull is calculated based on the energy of 86 competing phases imported from the Materials Project\cite{Jain/PingOng/Hautier:2013} (listed in the supplemental information). DFT total energies (including those for competing phases) are calculated using a consistent setup (see Methods) employing HSE06 and PBEsol functionals for comparison. On the one hand, hybrid calculations are based on the energies of all dynamically stable phases resulting from the workflow presented in Fig.  \ref{fig_structuralSearch} (25 structures initialised from Glazer tilts and 25 from Random search). On the other hand, the starting structures with Glazer tilt (25 structures) are added for PBEsol evaluation. The results are presented in Fig. \ref{fig_Thermodyn}.

Starting from the lowest-energy tilts (orange diamonds) and looking for dynamical stability, we can see that while imposing general tilts patterns lowered the energy (Fig. \ref{fig_tilts}e) significantly, most of the structures are dynamically unstable and result after phonon mapping in significantly lower energy structures going from a few meV/atom (e.g. TlWN$_3$) to almost 200 meV/atom (e.g. SnNbN$_3$). Alternatively, the structures initialised with random positions of the atoms result in being energetically more favourable. Nevertheless, most materials have a relatively small energy difference between their polymorphs, which is even more veracious for those below 200 meV/atom above the hull. This suggests that appropriate growth conditions (e.g. strain engineering or temperature) might allow the stabilisation of metastable phases with perovskite structure. The exact energy numbers are reported in Table \ref{tab_abc_E_P}. Considering our PBEsol results, we obtain 17 materials with an energy of 200 meV/atom or less above the hull, with six materials sitting directly on the hull. 

\begin{figure}[h]
\includegraphics[width=\columnwidth]{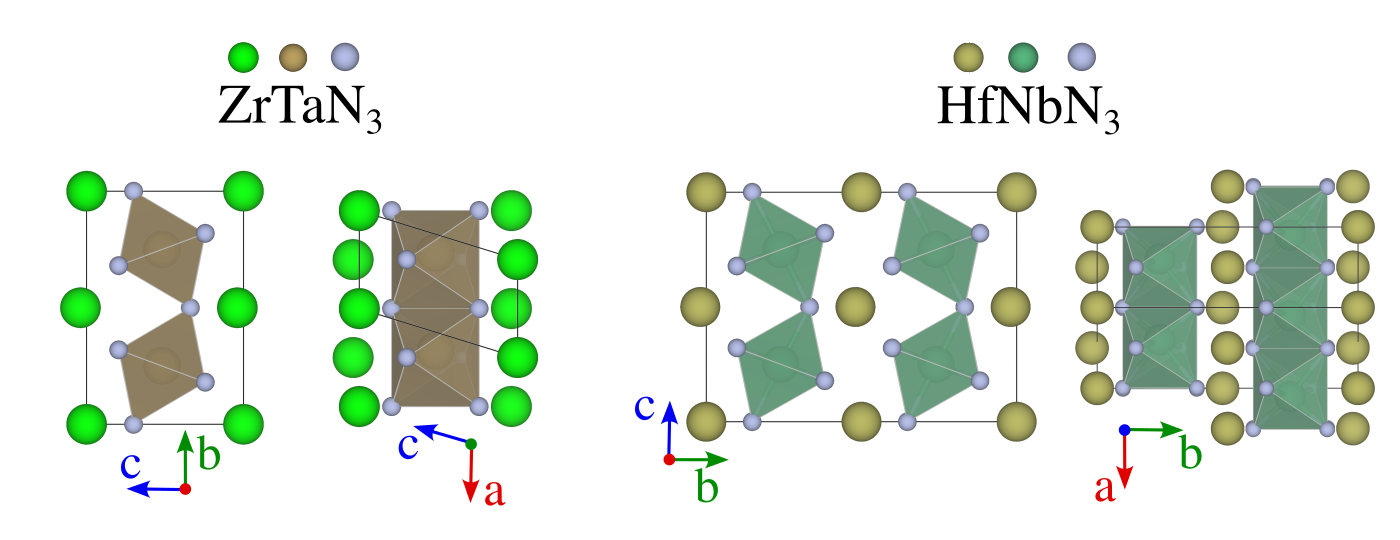}
\centering
\caption{Crystal structures of ZrTaN$_3$ and HfNbN$_3$ identified by random structure searches. While perovskites feature a connected network of BX$_6$ octahedra surrounded by A-cations, these structures are characterised by layers of A-cations alternated by layers of octahedra, which are corner shared along the vertical direction (left-hand-side of each material) and side shared along the perpendicular direction (right-hand-side of each material).}
\label{fig_layerstructure}
\end{figure}

Despite its extensive usage for predicting the thermodynamic stability of new nitrides, the PBEsol functionals display systematic errors for calculating the band gap, crucial for distinguishing between semiconductors and metals, as well as for calculating the enthalpy of formation of gas-phase N compounds, which results in the wrong estimation of the energy hull.\cite{UrregoOrtiz/Builes/Vallejo:2021} We evaluate this inaccuracy by recomputing the thermodynamic stability with HSE06 functional. From Fig. \ref{fig_Thermodyn}, it appears that PBEsol functionals overestimate the stability of the nitrides: out of the 17 materials with a distance of 200 meV/atom or lower to the hull, 12 remain within this range when more accurately computing the energies. Finally, we highlight that eight compounds adopt a perovskite-like structure in their ground state (InWN$_3$, LaMoN$_3$, LaWN$_3$, SnNbN$_3$, SnSbN$_3$, SnTaN$_3$, TlWN$_3$ and TlWN$_3$). The others adopt a layered-type structure shown in Fig. \ref{fig_layerstructure} for two prototype materials.

\subsection{Evaluating the structural search} \label{sec:literature_comparison}
In Fig. \ref{fig_Thermodyn}, we report with black dots the energy for six materials predicted before the current study and compare them with our findings. Lanthanum tungsten nitride (LaWN$_3$) was first predicted to adopt a structure with $R3c$ symmetry in its ground state,\cite{SarmientoPerez/Cerqueira/Korbel:2015} which was recently confirmed through synthesis.\cite{Talley/Perkins/Diercks:2021} Following our structural search workflow (Fig. \ref{fig_structuralSearch}) we obtain two polymorphs with $Pc$ symmetry (A) and $Cc$ symmetry (G). While differing in symmetry, both of our polymorphs are very close in energy (less than 2 meV/atom with PBEsol) to the true ground state and possess the $a^-a^-a^-$ tilt pattern characteristic of the $R3c$ space group. We use ISODISTORT\cite{Stokes/Hatch/Campbell} to decompose each structure into irreducible modes, using the $Pm\bar{3}m$ as parent structure, and find that both our polymorphs contain the $\Gamma$ (polar) and $R$ (tilts) modes characteristic from the $R3c$ ground state\cite{Fang/Fisher/Kuwabara:2017} as the main contributors to the distortions present in the crystal structures. The difference in symmetry results from residual distortion modes, suggesting that LaWN$_3$ has several low-energy metastable polymorphs.

LaMoN$_3$ was predicted to adopt a non-perovskite structure with $C2/c$ symmetry.\cite{SarmientoPerez/Cerqueira/Korbel:2015} Using our settings, we find that LaMoN$_3$ in this phase stands at around 70 meV/atom above the hull (PBEsol). We identify this material's ground state as a perovskite structure with $R3c$ symmetry, identified as the lowest energy phase in both our workflows, and standing on the hull using PBEsol, and at only 25 meV/atom above the hull with hybrid functional. In Table \ref{tab_abc_E_P}, we report a $Pc$ symmetry for one of the workflows for consistency with the tolerance to detect symmetry. However, the $R3c$ phase is also found when lowering the tolerance for symmetry detection of the Glazer structure. Moreover, the energy degeneracy, the almost similar lattice parameters, and the density of states (see supplementary Fig. S10) show the similarity of both structures.

\setlength{\tabcolsep}{10pt}
\renewcommand{\arraystretch}{1.2}
\begin{table*}[!htbp]
  \centering
\resizebox{\textwidth}{!}{%
  \begin{tabular}{ l c c c c c c c c c c c}
  \toprule[1.5pt]
   & &
  \multicolumn{3}{c}{\textbf{Energies}} &
  \multicolumn{3}{c}{\textbf{Lattice parameters}} &
  \multicolumn{2}{c}{\textbf{Structure}}&
  \multicolumn{2}{c}{\textbf{Polarisation}} \\ 
  
  \cmidrule[0.7pt](lr){3-5}
  \cmidrule[0.7pt](lr){6-8}
  \cmidrule[0.7pt](lr){9-10}
  \cmidrule[0.7pt](lr){11-12}

 & & E$_{hull}$ & E$^I_{g}$ & E$^D_{g}$& a & b & c  & Spg & Type & P & E$_{f}$ \\
   \textbf{Material}& & [meV/atom] & [eV] & [eV] & [\r{A}] & [\r{A}] & [\r{A}]  &  &  & [$\mu C/cm^2$] & [kV/cm] \\
   \cmidrule[1.2pt](lr){1-12}
LaWN$_3$ & (A) & 0 & 2.04 & 2.28 & \emph{7.96} & 5.67 & 5.63 & 7 & Perovskite & $P_a = 23, P_c = 37$ & 1.5\\ 
 & (G) & 0 & 2.05 & 2.32 & 5.59 & 5.59 & \emph{8.08} & 9 & Perovskite & $P_a = 28, P_c = 26$ & 1.8 \\ 
 
YMoN$_3$ & (A)& 0 & 2.18 & 2.27 & 5.65 & 5.36 & 7.76 & 7 & Perovskite & $P_c = 49$ & 11.1 \\ 
 & (G)& 3 & 2.22 & 2.30 & \emph{7.59} & 5.68 & 5.50 & 31 & Perovskite & $P_c = 68$ & 6.9 \\ 
 
ZrTaN$_3$ & (A)& 4 & 0.45 & 1.06 & 3.04 & \emph{7.35} & 5.15 & 4 & Layered & metallic & – \\ 
 & (G)& 34 & – & 2.13 & 7.69 & \emph{7.71} & 10.50 & 14 & Perovskite & – & – \\ 
 
YWN$_3$ & (A)& 16 & 2.25 & 2.53 & \emph{7.74} & 5.63 & 5.45 & 31 & Perovskite & $P_c = 34$ & 16.3 \\  
 & (G)& 16 & 2.25 & 2.53 & \emph{7.74} & 5.63 & 5.45 & 31 & Perovskite & $P_c = 34$ & 16.3 \\ 
 
 LaMoN$_3$ & (A)& 25 & 2.08 & 2.44 & 5.64 & 5.64 & \emph{13.8} & 161 & Perovskite & $P_c = 66$ & 1.2\\ 
 & (G)& 25 & 2.05 & 2.29 & \emph{7.94} & 5.64 & 5.63 & 7 & Perovskite & $P_a = 55, P_c = 38$ & 0.9 \\ 
 
HfTaN$_3$ & (A)& 29 & 0.55 & 1.14 & 3.02 & \emph{7.33} & 5.12 & 4 & Layered & metallic & – \\ 
 & (G)& 57 & – & 2.48 & 7.61 & 7.74 & 10.42 & 14 & Perovskite & – & – \\ 
 
 ZrNbN$_3$ & (A)& 40 & 0.74 & 1.27 & 3.06 & \emph{7.34} & 5.19 & 4 & Layered & metallic & – \\ 
 & (G)& 94 & – & 2.22 & 7.63 & \emph{7.87} & 10.49  & 14 & Perovskite & – & – \\ 

HfNbN$_3$ & (A)& 67 & – & 1.06 & 9.86 & \emph{7.32} & 3.04 & 13 & Layered & metallic & –\\ 
 & (G)& 114 & – & 2.34 & 7.50 & \emph{7.97} & 10.41 & 14 & Perovskite & – & – \\ 

ScWN$_3$ & (A)& 81 & 2.19 & 2.23 & 3.01 & \emph{7.30} & 10.20  & 13 & Layered & – & –\\ 
 & (G)& 100 & 2.56 & 2.64 &  \emph{7.57} & 5.54 & 5.21 & 31 & Perovskite & $P_c = 28$ & 50.4 \\ 

ScMoN$_3$ & (A)& 124 & 2.07 & 2.10 & \emph{7.18} & 3.03 & 9.69 & 2 & Layered & – & – \\ 
 & (G)& 168 & 2.02 & 2.26 & 5.56 & 5.16 & \emph{7.54} & 33 & Perovskite & $P_c = 39$ & 29.7\\ 

ZrVN$_3$ & (A)& 123 & 1.41 & 1.66 & 3.02 & \emph{7.03} & 9.50 & 19 & Layered & metallic & – \\ 
 & (G)& 153 & 2.25 & 2.35 & \emph{14.56} & 5.10 & 5.55 & 31 & Perovskite & $P_c = 1.7$ & 872.7\\ 

HfVN$_3$ & (A)& 144 & 1.53 & 1.57 & 2.99 & 9.44 & \emph{7.62} & 7 & Layered & metallic & – \\ 
 & (G)& 176 & 2.20 & 2.27 & \emph{7.25} & 5.51 & 5.06 & 31 & Perovskite & $P_c = 64$ & 28.9\\

  \bottomrule[1.5pt]
  \end{tabular}
   }
  \caption{Characteristics of the 12 candidates below 200 meV/atom above the hull for HSE06 energy calculations. We report the information about each material's AIRSS (A) and Glazer (G) structures. The \textbf{Energies} are calculate with HSE06 functional and provide the energy above the hull (E$_{hull}$), indirect (E$^I_{g}$) and direct (E$^D_{g}$) band gaps. The \textbf{Lattice parameters} are obtained from the conventional unit cell relaxed with PBEsol functional. The value in italics indicates the axis along the cartesian $z$ axis that would correspond to the out-of-plane direction for a thin film. Below \textbf{Structure} we report the space group and the type of structure (Perovskite/Layered). Finally, under \textbf{Polarisation}, we report the non-null components of the polarisation vector projected on each direction of the conventional unit cell and the corresponding value of the electric field required to switch the polarisation from +P to -P (see Methods). Note that the field is assumed to be applied along the direction with the largest polarisation component. The values "metallic" refer to metallic structures relaxed with PBEsol.}
  \label{tab_abc_E_P}
\end{table*}

YMoN$_3$ and YWN$_3$ have been reported to adopt the same structure as LaMoN$_3$ (space group 15)\cite{SarmientoPerez/Cerqueira/Korbel:2015} and we find that they are respectively at about 25 meV/atom and 125 meV/atom above the hull (PBEsol calculations) in these reported phases. In the present work, YMoN$_3$ is found to be a promising new nitride perovskite candidate. Indeed, in its ground state, this material stands on the hull with a structure of $Pc$ symmetry (A). We identify a polymorph with $Pmn2_1$ symmetry that is only 3.5 meV/atom higher in energy. In the case of YWN$_3$, the ground state adopts a perovskite structure with $Pmn2_1$ symmetry (A and G) and stands at only a few meV/atom above the hull.

Finally, ScMoN$_3$ and ScWN$_3$ are mentioned in literature as unlikely to be stable in a perovskite structure.\cite{Sherbondy/Smaha/Bartel:2022} While no information about the structure could be found,\cite{Sherbondy/Smaha/Bartel:2022} we take as reference the structure with $P\bar{1}$ symmetry found on Materials Project\footnote{mp1246879 and mp1246967} and find that their energies are about 200 meV/atom far from the hull. We find that both materials are dynamically stable in a perovskite structure, respectively, with $Pna2_1$ symmetry for ScMoN$_3$ and $Pmn2_1$ for ScWN$_3$. Nevertheless, the ground states are non-perovskite structures with alternated Sc and Mo (W respectively) layers and $P\bar{1}$ symmetry for ScMoN$_3$, and $P2_1/c$ for ScWN$_3$. Note that the energy difference between the polymorphs of the latter material is less than 20 meV/atom.

While some compositions were already predicted before our work, our systematic structural search workflow is proved robust in identifying ground-state structures. Indeed, it allowed the identification of what is likely the ground state for already identified and new candidates. The following section provides a deeper analysis of the most promising materials.

\subsection{Promising candidates} \label{sec:promising_candidates}
From the 12 candidates in Table \ref{tab_abc_E_P}, only HfNbN$_3$ is centrosymmetric in both identified structures ((A) and (G)). All other candidates have a non-centrosymmetric structure allowing by symmetry the existence of a spontaneous polarisation and possibly ferroelectricity in these materials. As a reminder, a ferroelectric material must be non-centrosymmetric in its ground state, with a non-zero spontaneous polarisation whose orientation can be inverted by applying an electric field. We evaluate the first two criteria by calculating the polarisation (see Methods). As presented in Table \ref{tab_abc_E_P}, all the polar candidates have a significant polarisation along at least one cartesian direction. In particular, we find that YMoN$_3$ exhibits a polarisation of $49 ~\mu C/cm^2$ in its ground state (standing on the hull) and $68 ~\mu C/cm^2$ in a metastable phase, YWN$_3$ has a polarisation of $34 ~\mu C/cm^2$ and LaMoN$_3$ while adopting a similar structure than LaWN$_3$ has polarisation of $66 ~\mu C/cm^2$ along the [111] pseudo-cubic direction. These values are comparable to the best oxides (BiFeO$_3$ or PbTiO$_3$) and the prototypical polar nitride (LaWN$_3$). 

\begin{figure*}[t]\
\includegraphics[width=0.95\linewidth]{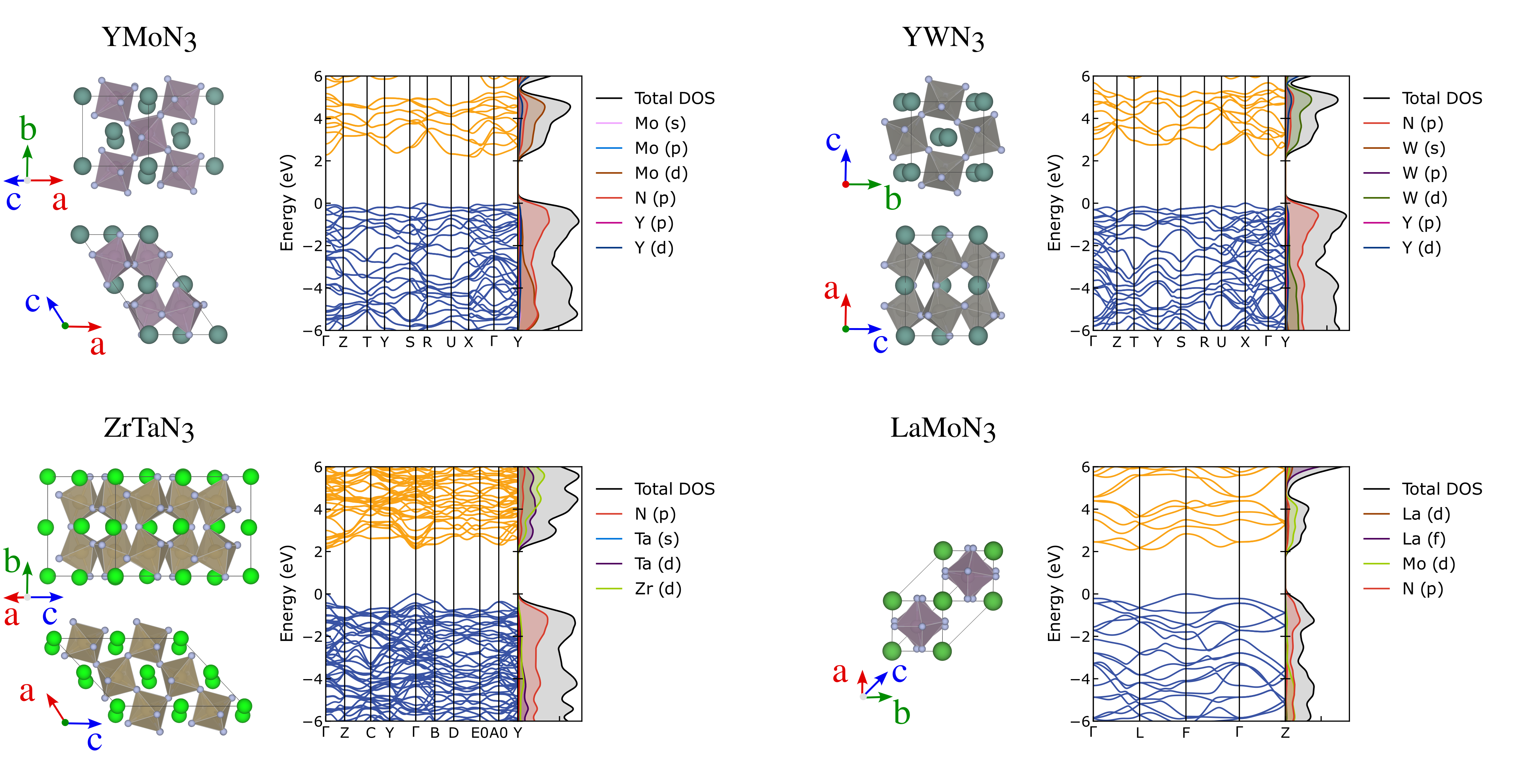}
\centering
\caption{Crystal structure, electronic band structures and projected density of states of YMoN$_3$ (A), ZrTaN$_3$ (G), YWN$_3$ and LaMoN$_3$ (A).}
\label{fig_materials}
\end{figure*}

Next, to evaluate the potential for being ferroelectric, we estimate the electric field required to switch the polarisation (see Methods for details on the calculations). All polar materials, except HfVN$_3$, can potentially be switched with an appropriate electric field of the order of a few kv/cm. While this estimation is undoubtedly not highly rigorous, it reasonably estimates the electric field intensity required to switch the polarisation. It shows that this is likely to happen before the material breaks down.

ZrTaN$_3$ (A), HfTaN$_3$ (A) and ZrNbN$_3$ (A) are found to be metallic with PBEsol calculations, and their polarisation could, therefore, not be calculated. Using hybrid functional, they exhibit a small band gap of 0.45 eV, 0.55 eV and 0.74 eV, for ZrTaN$_3$, HfTaN$_3$ and ZrNbN$_3$ respectively. This indicates that while their pristine structures are not metallic, doping could, in principle, trigger metallicity. Those systems would therefore require further investigation to evaluate their potential as polar metals. While HfNbN$_3$ (A) and ZrVN$_3$ (A) are metallic, their structure is centrosymmetric (not polar).

Considering their distance of 25 meV/atom – corresponding to the thermal energy at room temperature – or less to the hull, their significant polarisation value and their likelihood of being switchable, we present YMoN$_3$, YWN$_3$ and LaMoN$_3$ as new ferroelectric materials very likely synthesisable. On the other hand, while ScWN$_3$ (G) and HfVN$_3$ (G) are promising in polarisation and switching fields, their energy above the hull is significantly higher and, therefore, more challenging. Finally, ZrVN$_3$ (G) is the least promising candidate for ferroelectricity due to its high energy above the hull and large switching field.

Further, than ferroelectricity, YMoN$_3$ (both structures) and ZrTaN$_3$ (G) seem promising for thermoelectric applications due to their electronic band structure presenting a mix of flat non-disperse bands with large charge career effective mass, related to high Seebeck, and disperse band edges with high career mobility, linked to high-electrical conductivity. These materials' crystal and electronic band structures are presented in Fig. \ref{fig_materials} along with those of YWN$_3$ and LaMoN$_3$ for comparison. The electronic and crystal structures of the 25 most promising candidates can be found in the supplementary in Fig S9-S13.

YMoN$_3$ (A) has an indirect band gap of 2.17 eV with the valence band maximum between (VBM) $S$ and $R$, and the conduction band minimum (CBM) between $X$ and $\Gamma$. The valence band has hole effective masses of 0.71 $m_e$ and 0.93 $m_e$ along VBM-$S$ and VBM-$R$, respectively. The conduction band has electron effective masses of 0.42 $m_e$ and 1.14 $m_e$ along CBM-$X$ and CBM-$\Gamma$, respectively. ZrTaN$_3$ (G) has a direct band gap of 2.13 eV at $\Gamma$ with hole effective masses of 0.49 $m_e$ along $\Gamma-Y$ and 0.57 $m_e$ along $\Gamma-B$, and electron effective masses of 0.34 $m_e$ and 0.40 $m_e$ along $\Gamma-Y$ and $\Gamma-B$, respectively. Both materials exhibit low hole and electron effective masses, which is promising for high conductivity, assuming that the materials are both p-type and n-type dopable.

\section{Conclusion and summary} \label{sec:conclusion}
We presented a DFT-based high throughput study to identify nitride perovskite materials. Contrary to previous works, our approach filtered candidate compositions based on chemical and electrostatic criteria only, without prior knowledge of the crystal structure. The latter was thoroughly investigated by actively enumerating possible octahedral tilt systems combined with global optimisation. Starting from 3660 ABN$_3$ compositions, our workflow identified 279 chemically feasible candidates, and 25 were further investigated. We identified 12 dynamically and thermodynamically stable nitride perovskites materials and propose that five of them (LaWN$_3$, YMoN$_3$, YWN$_3$, ZrTaN$_3$ and LaMoN$_3$) are particularly promising for ferroelectric memory devices. 

From a methodological point of view, our work shows the limitations of traditional high throughput studies relying on common structures that, in some cases, can fail in identifying the correct ground state structure (e.g. LaMoN$_3$). While the computational cost of thoroughly exploring the potential energy surface is high, our mixed crystal structure search approach combining symmetry guided distortions (octahedral tilts) complemented by crystal structure prediction is robust in finding low-energy structures. 

Our work has implications for future experimental and theoretical studies. On the one hand we provided highly reliable studies of the stability of 25 closed-shell materials, which we hope will serve as starting point for future synthesis. On the other hand, we uncovered a wide chemical space of accessible new nitride materials using chemically based criteria. While the current study investigated the thermodynamic stability of 25 candidates out of the 279 predicted systems, their chemical feasibility should be enough to motivate further theoretical studies. We focused here on the materials with closed-shell. Still, we anticipate that other of our candidates with open-shell might display richer chemistry and exhibit electric and magnetic polarisation, opening the way to other technologically relevant classes of materials such as multiferroics.

\section{Methods} \label{sec:methods}
\subsection{Electronic structure calculations}
First-principles calculations were performed using Kohn-Sham DFT \cite{Kohn/Sham:1965} with the projector augmented wave (PAW) method \cite{Blochl:1994} as implemented in the Viena \textit{ab initio} simulation package (VASP). \cite{Kresse/Furthmuller:1996,Vasp3:1996} We used the PBEsol functional form of the generalized gradient approximation (GGA) \cite{PBEsol:2008,Perdew/Burke/Ernzerhof:1996} for structural relaxations, with a Hubbard U correction for the transition metals listed in the Materials Project \cite{Jain/PingOng/Hautier:2013}, adopting the recommended $U_{eff}$ values, optimal for oxides and halides. For the relaxations, the forces on each atom were minimised to below 0.01 eV/Å. The kinetic energy cutoff and the k-point grid were determined individually for each material considered (including the competing phases). The corresponding values can be found in the Supporting Information. 
While the structure can be accurately resolved by PBEsol functional, the electronic bandgap tends to be underestimated. Furthermore, the Hubbard U values recommended by Materials Project are tailored for oxides but not nitrides, which can introduce discrepancies in computing the formation energy of specific compositions. To overcome these issues, the structures were relaxed again. The corresponding electronic structure and the thermodynamic stability were computed with the hybrid Heyd–Scuseria–Ernzerhof functional HSE06, which includes 25\% of Hartree-Fock exact exchange.\cite{Heyd/Scuseria:2004,Heyd/Peralta/Scuseria:2005} The converged parameters determined for PBEsol calculations were kept for the hybrid ones.

\subsection{Random Structure Search}

Ab initio random structure searching (AIRSS)\cite{AIRSS1:Pickard/Needs:2006,AIRSS2:Pickard/Needs:2011} is used to find and validate low energy structures for perovskite compositions of the interests. The initial random structures generated contains two or four formula units of ABN$_3$. Since the searches aim to validate the predicted structures, they are non-exhaustive. For most compositions, only 200 generated structures have been relaxed containing two formula units, and 400 structures for four formula units. The species-wise minimum separations are used to bias the random structure generation process using the pair-wise distances between A, B, and N sites from a cubic perovskite cell of lattice constant 3.98 \AA, and the target volume of the generated structures is set to 75 \AA$^3$. This provides a more robust test of the likelihood of the perovskite structure.

For searching, density functional theory calculations are carried out using CASTEP\cite{CASTEP:2005} with a plane wave cut-off energy of 340 eV, reciprocal space sampling spacing of 0.07 2$\pi$ \AA$^{-1}$, and the PBEsol exchange-correlation functional.\cite{PBEsol:2008} The on-the-fly generated core-corrected ultrasoft pseudopotentials are used from the CASTEP built-in QC5 potential library. The DISP\cite{Zhu:2022} package is used for data management as workflow automation. 

For each material, the ten lowest-energy structures were then relaxed with VASP and coherent parameters to select the lowest-energy structure and evaluate its dynamical stability.

\subsection{Dynamical stability and phonons}
To verify that the identified crystal structures are dynamically stable, phonon frequencies are calculated at selected symmetry points using the frozen-phonon method, as implemented in the PHONOPY package. \cite{Togo/Tanaka:2015} Whenever an imaginary frequency is observed, the corresponding eigenvector's distortions are frozen using ModeMap,\cite{Skelton/Burton/Parker:2016} and the newly obtained structure is fully relaxed. \cite{} When several imaginary frequencies are obtained, the described procedure is executed individually for the lowest imaginary frequency at each high-symmetry point. The lowest-energy resulting structure is selected to re-evaluate dynamical stability. This procedure, previously applied with success for other families of materials,\cite{Rahim/Skelton/Savory:2020,Grosso/Spaldin:2021} is repeated until a structure containing only real frequencies over the Brillouin zone is identified.

\subsection{Polarisation and switching field}
The polarisation is estimated in the conventional unit cell by multiplying the displacement of each atom concerning a centrosymmetric higher-symmetry structure and its corresponding Born Effective Charges tensor (BEC). The result is then normalised by the volume of the unit cell and expressed in [$\mu C/cm^2$]. To construct the reference centrosymmetric structure, we invert the coordinates of each atom in the unit cell 
\begin{center}
$[x,y,z] \rightarrow [-x,-y,-z]$
\end{center}
and create the structure obtained by interpolating halfway in between both structures. 

To evaluate the switchability of the polarisation we calculate the required energy to overcome the barrier separating the initial structure and its version with inverted coordinates. The nudged elastic band (NEB),\cite{Mills/Jonsson/Schenter:1995} as implemented in VASP, is used to find the minimum energy path. A set of 5 intermediate states was generated by linear interpolation between the endpoint structures, with fixed volume along the path. The NEB calculations were run until forces reached a convergence of 0.01 eV/\AA.

Finally, we approximate the required electric field to overcome the energy barrier by making use of the energy produced by the coupling of the electric field ($\vec{E}$) and polarisation ($\vec{P}$): 

\begin{center}
$\Delta E = - \vec{P} \cdot \vec{E}$
\end{center}

We approximate the necessary electric field by inverting the previous relation with $\vec{P}$ taken as the polarisation value in the initial structure and $\Delta E$ the energy barrier.

All these calculations are done with PBEsol.

\subsection{Electronic band structure and effective masses}
The electronic band structures and density of states are plotted using Sumo\cite{Ganose/Jackson/Scanlon:2018} and the effective masses were calculated using effmass.\cite{Whalley_effmass:2018}

\section*{Acknowledgments}
The authors acknowledge the use of the UCL Kathleen and Myriad High Performance Computing Facility (Kathleen@UCL, Myriad@UCL) and associated support services, in the completion of this work, and are grateful to the UK Materials and Molecular Modelling Hub for computational resources, which is partially funded by EPSRC (EP/P020194/1 and EP/T022213/1). Via our UK HEC Materials Chemistry Consortium membership, funded by the UK Engineering and Physical Sciences Research Council (EP/ R029431), this work used the ARCHER2 UK National Super- computing Service (https://www.archer2.ac.uk).

\bibliography{Bibliography}

\end{document}